\documentclass[10pt,a4paper]{article}
\usepackage{authblk}
\usepackage[utf8]{inputenc}
\usepackage{setspace}
\usepackage[margin=0.75in]{geometry}
\usepackage{graphicx}
\graphicspath{ {./figures/} }
\usepackage{subcaption}
\usepackage{amssymb}
\usepackage{amsthm}
\usepackage{amsmath}
\newtheorem{definition}{Definition}
\newtheorem{proposition}{Proposition}
\newtheorem{lemma}{Lemma}

\usepackage{xcolor}
\usepackage[authoryear,round]{natbib}

\title{Equilibrium Thresholds in the Repeated Prisoner's Dilemma with State Costs}
\author[1]{Omar Muhammad}
\author[1]{Amrit Mahendra Joseph}
\affil[1]{Indian Institute of Science}
\date{}

\begin{document}

\maketitle

\begin{abstract}
    Cooperation in repeated strategic interactions can depend on strategies that remember past behaviour and respond to deviations. When implementing these strategies carries a recurring complexity cost, players may benefit from adopting simpler implementations, potentially undermining cooperation. We study this trade-off in the infinitely repeated Prisoner's Dilemma by modelling strategies as deterministic finite automata. Each player's utility is their long-run average payoff minus a recurring maintenance cost proportional to the number of automaton states. We study four cost thresholds governing the equilibrium payoffs that remain attainable as state costs increase. We characterize the threshold for eventual mutual cooperation in Nash equilibrium and show that every positive state cost removes some payoffs attainable when states are free. We also establish bounds on the thresholds beyond which mutual defection is the only equilibrium outcome under Nash and semi-perfect equilibrium, and conjecture their exact values.
\end{abstract}

\section{Introduction and Motivation} 

The standard analysis of repeated games often abstracts away the resources required to implement a strategy. Bounded rationality brings these limitations into the analysis: remembering past behaviour and responding to deviations requires memory and computation, whose cost may affect a player's choice of strategy.

We model strategies as Moore machines: deterministic finite automata with an action assigned to each state. Transitions update the state after observing the opponent's action. Charging for each state makes players weigh the benefits of cooperation against the cost of implementing the strategy that sustains it.

This tradeoff can undermine familiar cooperative strategies. Consider two players using grim trigger: each cooperates initially and switches permanently to defection if the other defects. Each strategy requires a cooperative state and a punishment state. When the two machines interact, neither punishment state is visited. Either player can instead use a one-state machine that always cooperates, obtaining the same payoff against the opponent while paying less for maintenance. Thus, although the original machines produce mutual cooperation, they do not form an equilibrium when states have a positive cost.

We study this trade-off in the infinitely repeated Prisoner's Dilemma through four thresholds: the loss of the full equilibrium payoff set, the loss of eventual mutual cooperation, and the disappearance of non-defection outcomes under Nash and semi-perfect equilibrium.

Our aim is to determine how much state cost different equilibrium behaviours can withstand. The distinction between Nash and semi-perfect equilibrium matters because a machine that is worth maintaining at the start may become unnecessarily large after its introductory phase.

\section{Related Work}
Folk theorem results show how future punishment can sustain a wide range of equilibrium payoffs (any feasible and individually rational vector) including cooperation in repeated games when there is no charge for complexity~\citep{friedman1971non}. \citet{aumann1981survey} proposed finite automata for modelling strategies, since the infinite memory required for many standard supergame strategies is physically unrealizable and computationally unrealistic. \citet{neyman1985bounded} showed that restrictions on strategy complexity can support cooperation even in the finitely repeated Prisoner's Dilemma, despite the usual backward-induction argument. This was extended in \citet{megiddo1986play} showing that the sustainability of cooperation persists even when strategies are modelled by size-restricted Turing Machines.

\citet{rubinstein1986finite} and \citet{abreu1988structure}
study strategy complexity as a cost.
\citet{rubinstein1986finite} uses lexicographic preferences:
players maximize average payoff first and minimize states second. This can destabilize strategies with unused punishment states,
which prescribe adverse responses to an opponent's deviation:
a player can obtain the same payoff against the opponent
using a smaller machine.
However, this does not rule out all Nash equilibria with
eventual mutual cooperation. Rubinstein's stronger
semi-perfect equilibrium requirement rules out mutual
cooperation in the eventual cycle, while permitting mutual
defection and certain periodic exploitation cycles. \citet{bankssundaram1990} include transition structure in the complexity measure and obtain stage-game Nash play in every period.

Under a modified evolutionary stability criterion with lexicographic costs, \citet{binmore1990} require outcomes to maximize the sum of the players' payoffs. \citet{cooper1996evo} instead subtracts a linear state cost from time-average payoffs and obtains a
folk-theorem-type result: every payoff between mutual defection and mutual cooperation can be approximated by a neutrally stable strategy for sufficiently small positive costs. \citet{veelen2019} extend this result to discounting and establish a uniform invasion barrier for the one-state defector. These results distinguish the existence of cooperative
equilibria from their evolutionary stability. Our focus is on how the existence of Nash and semi-perfect
equilibria changes as the per-state cost increases.

\section{Model and Equilibrium Concepts}

Let $G=\langle S_1,S_2,u_1,u_2\rangle$ be a two-player normal-form game where $S_i$ and $u_i:S_1\times S_2\to\mathbb{R}$ are respectively a finite set of actions and the payoff function for player $i$. Its supergame repeats $G$ at periods $t=1,2,\ldots$, with each player observing the opponent's action after each period. A supergame strategy determines a player's choice of action at period $t$ as a function of the previous $t-1$ outcomes.

We restrict to strategies implemented by Moore machine
$M_i=\langle Q_i,q_i^0,\sigma_i,\mu_i\rangle$, where
$Q_i$ is a finite state set, $q_i^0\in Q_i$ is the initial
state, $\sigma_i:Q_i\to S_i$ is the output function, and
$\mu_i:Q_i\times S_{3-i}\to Q_i$ is the transition function.
At period $t$, player $i$ in state $q_i^t$ plays
$s_i^t=\sigma_i(q_i^t)$ and moves to
$q_i^{t+1}=\mu_i(q_i^t,s_{3-i}^t)$.
Write $|M_i|=|Q_i|$, and let $M_i(q)$ denote the same
machine started at $q$, retaining its full state set.
The joint run $(q_1^t,q_2^t)$ is deterministic on the finite
product $Q_1\times Q_2$, hence ultimately periodic.
We define player $i$'s supergame payoff $\pi_i(M_1,M_2)$
as their average payoff over the eventual joint cycle,
equivalently the long-run average of the generated payoffs.

For a per-state maintenance cost $\lambda\ge0$, we define utility for a player as $U_i(M_1,M_2)=\pi_i(M_1,M_2)-\lambda|M_i|$.

\begin{definition}
A pair $(M_1,M_2)$ is a Nash equilibrium if neither player
can increase utility by changing their machine at the start, holding the opponent's machine fixed:
$U_1(M_1,M_2)\ge U_1(M_1',M_2)$ and
$U_2(M_1,M_2)\ge U_2(M_1,M_2')$
for all finite replacement machines $M_1',M_2'$.
\end{definition}

\begin{definition}
A pair $(M_1,M_2)$ is a semi-perfect equilibrium (SPE) if neither player can increase utility by replacing their
machine at any period $t\ge1$ along the induced play,
while the opponent continues from its current state:
$U_1(M_1(q_1^t),M_2(q_2^t))
\ge U_1(M_1',M_2(q_2^t))$ and
$U_2(M_1(q_1^t),M_2(q_2^t))
\ge U_2(M_1(q_1^t),M_2')$
for all finite replacement machines $M_1',M_2'$.
\end{definition}

Every SPE is an NE. We specialize to the Prisoner's Dilemma, where each
player chooses whether to cooperate ($C$) or defect ($D$):
\[
\begin{array}{c|cc}
 & C & D \\\hline
C & R,R & S,T \\
D & T,S & P,P
\end{array}
\]
Each entry lists the row and column players' payoffs. Here $R$ is the reward for mutual cooperation, $P$ the
punishment payoff for mutual defection, $T$ the temptation
payoff for defecting against a cooperator, and $S$ the
sucker's payoff for cooperating against a defector.
We assume $T>R>P>S$ and $2R>T+S$: each player individually benefits from defection regardless
of the other's action, yet mutual cooperation uniquely
maximizes their combined payoff.
Let $M_C$ and $M_D$ be the one-state machines that always
cooperate and always defect, respectively.

\section{Threshold Results}
We state the main results and their proofs.
A payoff profile is the pair
$v=(\pi_1(M_1,M_2),\pi_2(M_1,M_2))$ of players' game
payoffs, before subtracting state costs.
% Setup additions to make before submission: |M_i|:=|Q_i|; M_i(q) is M_i restarted at q with its full state set retained; M_D and M_C are the one-state always-defect and always-cooperate machines. SPE tests every continuation along the generated play, including the introductory part, not every possible history.

\subsection{From free states to positive costs}
\begin{proposition}[No costs for states]\label{prop:zero}
In the repeated Prisoner's Dilemma with $\lambda=0$, a payoff profile $v=(v_1,v_2)$ is a Nash equilibrium payoff of the machine game if and only if $v$ is a rational convex combination of $(R,R),(S,T),(T,S),(P,P)$ and $v_i\ge P$ for $i\in\{1,2\}$. The same characterization holds for semi-perfect equilibrium payoffs.
\end{proposition}
\begin{proof}
Every finite-machine pair generates an eventual cycle. The frequencies of its four outcomes are rational, so its payoff is a rational convex combination of the four stage-game payoff vectors. A deviation to $M_D$ guarantees at least $P$, which proves necessity for NE and hence for SPE.

Conversely, choose a finite sequence of outcome pairs with the prescribed rational frequencies. Give each player a machine that repeats their component of this sequence, using one state for each position, and an additional absorbing state that defects. At a sequence state the machine advances if the opponent plays the prescribed action; otherwise it enters the absorbing state.

If both players follow the sequence, their payoffs are $v_i$. Against either prescribed machine, a replacement either follows the sequence forever or first departs from it at a finite time. In the latter case the opponent defects forever thereafter, and the deviator's average payoff is at most $P\le v_i$. In the former case the payoff remains $v_i$. Since states are free, reducing machine size gives no utility improvement. This argument applies from every position of the prescribed cycle, proving SPE as well as NE.
\end{proof}

% Now consider infitesimal state costs. This can be thought of as lexicographic preferences.
Lexicographic preferences prioritize game payoff and use machine size to break ties. Lemma~\ref{lem:structure} below establishes that every positive-cost equilibrium satisfies the corresponding lexicographic equilibrium conditions. Under this setup, unused punishment states cannot be retained merely to deter deviations. Rubinstein's semi-perfect equilibria exclude mutual cooperation from the eventual cycle, while allowing mutual defection and individually rational exploitation cycles~\citep{rubinstein1986finite}. Nash equilibrium can still support eventual mutual cooperation~\citep{abreu1988structure}.

Let $V^*$ be the set of payoff profiles characterized in Proposition~\ref{prop:zero}. We ask how much state maintenance can cost before particular
equilibrium outcomes become impossible.
The following suprema give the upper cost limits for
preserving the entire set $V^*$, sustaining mutual cooperation
in NE, and sustaining any payoff other than mutual defection
in SPE and NE, respectively:
\begin{align*}
\lambda_1&=\sup\{\lambda\ge0:\text{every }v\in V^*\text{ is an NE payoff}\},\\
\lambda_2&=\sup\{\lambda\ge0:\text{an NE yields }(R,R)\},\\
\lambda_3&=\sup\{\lambda\ge0:\text{an SPE yields }v\ne(P,P)\},\\
\lambda_4&=\sup\{\lambda\ge0:\text{an NE yields }v\ne(P,P)\}.
\end{align*}
Here $(R,R)$ means eventual mutual cooperation; a finite introductory phase does not affect the average payoff. The distinction between NE and SPE matters because a machine chosen at the start may cease to be optimal after its introductory states have been used.

\subsection{Best responses and existence of the thresholds}
\begin{proposition}[Upper bound and existence]\label{prop:upper}
For any state-maintenance cost $\lambda>R-P$, the unique Nash equilibrium of the machine game is $(M_D,M_D)$, yielding $(P,P)$. All four thresholds therefore exist, with
\[
0\le\lambda_1\le\lambda_2\le\lambda_4\le R-P,
\qquad 0\le\lambda_3\le\lambda_4.
\]
\end{proposition}

\begin{lemma}[{\citealp[Lemma~1]{abreu1988structure}}]\label{lem:fixed}
Against a fixed opponent $M_2$, a machine with at most $|M_2|$ states attains the maximum game payoff over all finite-machine responses.
\end{lemma}
\begin{proof}
Form a directed multigraph whose vertices are the states of $M_2$. For each state $q$ and action $a\in\{C,D\}$, add the edge
\[
q\longrightarrow\mu_2(q,a),\qquad
w(q,a)=u_1(a,\sigma_2(q)).
\]
Restrict attention to vertices reachable from the initial state. Any finite response generates an eventually periodic walk in this graph. Its eventual closed walk can be decomposed into simple cycles, and its mean weight is a weighted average of their means. It is therefore at most the maximum reachable simple-cycle mean $m^*$.

Choose a simple cycle attaining $m^*$ and a shortest path from the initial state to that cycle. The path can be taken simple and disjoint from the cycle except at its endpoint. A machine can follow the actions labelling this path and then repeat the actions labelling the cycle. One state per visited graph vertex suffices, so it uses at most $|M_2|$ states. Its average payoff is $m^*$, proving the claim.
\end{proof}

\begin{lemma}[Finite-cost and lexicographic equilibria]\label{lem:structure}
For $\lambda>0$, every NE has $|M_1|=|M_2|$, and each machine maximizes the game payoff against the other. Consequently, every finite-cost NE is a lexicographic NE, and every finite-cost SPE is a lexicographic SPE.
\end{lemma}
\begin{proof}
If $|M_1|>|M_2|$, Lemma~\ref{lem:fixed} supplies player 1 with a response having at most $|M_2|$ states and at least the current game payoff. At positive cost this strictly improves utility. Interchanging the players proves equal sizes, say $n$.

If either player could obtain a higher game payoff, Lemma~\ref{lem:fixed} would attain the maximum using at most $n$ states, again improving utility. Finally, an equal-payoff response with fewer states would also improve utility. Thus each player maximizes game payoff and, among responses attaining that payoff, minimizes states. These are exactly the lexicographic best-response conditions. For SPE, apply the argument to each continuation pair, retaining the full state sets as in the definition.
\end{proof}

\begin{lemma}[One-state machines]\label{lem:one}
Under strictly positive state costs: (i) a pair of one-state machines is an NE if and only if both are $M_D$; (ii) any NE with payoff $(P,P)$ uses $(M_D,M_D)$; and (iii) in every NE with payoff different from $(P,P)$, both machines have at least two states.
\end{lemma}
\begin{proof}
Against a constant cooperator, always defecting gives the largest possible game payoff $T$; against a constant defector, it gives the largest possible payoff $P$. In either case $M_D$ attains this payoff with the minimum possible size, one state. Any larger machine has strictly lower utility if it attains the same payoff, and cannot do better in game payoff. Thus an equilibrium involving a one-state machine must use $M_D$ for both players, proving (i) and (iii). For (ii), if a player earns $P$ with more than one state, switching to $M_D$ guarantees at least $P$ while reducing cost. Both machines in a $(P,P)$ equilibrium must therefore have one state.
\end{proof}

\begin{proof}[Proof of Proposition~\ref{prop:upper}]
The pair $(M_D,M_D)$ is an equilibrium for every $\lambda\ge0$. Consider any positive-cost equilibrium with a different payoff. By Lemma~\ref{lem:one}, both machines have at least two states. Since every stage outcome has total payoff at most $2R$,
\[
U_1+U_2\le2R-\lambda(|M_1|+|M_2|)\le2R-4\lambda.
\]
Each player can guarantee $P-\lambda$ with $M_D$, so equilibrium also requires $U_1+U_2\ge2P-2\lambda$. Combining these inequalities gives $\lambda\le R-P$. Lemma~\ref{lem:one}(ii) excludes any other machine pair yielding $(P,P)$ at positive cost.

At zero cost, Proposition~\ref{prop:zero} makes all four defining sets nonempty. Each is bounded above by $R-P$, so its supremum exists. Preserving all of $V^*$ entails preserving $(R,R)$, and $(R,R)\ne(P,P)$. Every SPE is an NE. These inclusions give the stated ordering.
\end{proof}

\subsection{The breakdown of the Folk Theorem: $\lambda_1$}
\begin{lemma}[State-cost bounds]\label{lem:bound}
For a profile with payoff $v_i$ and $|M_i|>1$, a necessary condition for NE is $\lambda\le(v_i-P)/(|M_i|-1)$. In particular, given $\lambda>0$, every machine in an NE satisfies
\[
|M_i|\le1+\left\lfloor\frac{T-P}{\lambda}\right\rfloor.
\]
\end{lemma}
\begin{proof}
In equilibrium, replacing $M_i$ by $M_D$ cannot improve utility. Since this replacement guarantees at least $P-\lambda$,
\[
v_i-\lambda|M_i|\ge P-\lambda.
\]
Rearranging gives the first bound when $|M_i|>1$. Using $v_i\le T$ and the integrality of $|M_i|$ gives the second. The quantity being charged is the player's machine size, not the length of the joint outcome cycle.
\end{proof}

\begin{proposition}[Immediate loss of the full payoff set]\label{prop:lambda1}
For any finite state cost $\lambda>0$, some payoff in $V^*$ ceases to be a Nash equilibrium payoff. In particular, $\lambda_1=0$.
\end{proposition}
\begin{proof}
For fixed $\lambda>0$, Lemma~\ref{lem:bound} gives a finite bound on each equilibrium machine's size. With two actions, there are only finitely many machines of bounded size up to state renaming, hence finitely many equilibrium payoff profiles. In contrast, $V^*$ contains infinitely many rational mixtures of $(R,R)$ and $(P,P)$. It cannot be attained in full at any positive cost. Proposition~\ref{prop:zero} shows that it is attained at zero cost, so $\lambda_1=0$.
\end{proof}

\subsection{Mutual cooperation in Nash equilibrium: $\lambda_2$}
Grim trigger (cooperates until the opponent defects, then
defects forever) fails at positive costs because its unused punishment state can be discarded. A punishment phase used at the beginning of play can avoid this deviation.

\begin{proposition}[Threshold for mutual cooperation]\label{prop:lambda2}
The maximum cost threshold for sustaining the mutual cooperation payoff $(R,R)$ in a Nash equilibrium is
\[
\lambda_2=\max_{N\in\mathbb Z_{>0}}g(N),
\qquad g(N)=\frac{N(R-P)-(T-R)}{N(N+1)}.
\]
\end{proposition}

\begin{proof}
Write $A=R-P>0$ and $B=T-R>0$. Then $g(N)=(AN-B)/(N(N+1))$ is positive for sufficiently large $N$ and tends to zero as $N\to\infty$. Its positive maximum is therefore attained at a finite integer.

\emph{Sufficiency.} Fix $N\ge1$ with $g(N)\ge0$. The password machine has states $q_1,\ldots,q_N,q_C$, starts at $q_1$, and outputs $D$ in $q_1,\ldots,q_N$ and $C$ in $q_C$. Define
\[
\mu(q_k,D)=q_{k+1}\ (k<N),\qquad
\mu(q_N,D)=q_C,\qquad \mu(q_C,C)=q_C.
\]
Every other transition returns to $q_1$. Two copies generate $(D,D)^N(C,C)^\infty$ and give each player utility $R-\lambda(N+1)$.

First bound the payoff of any response, irrespective of its size. In the opponent graph of Lemma~\ref{lem:fixed}, the cooperative self-loop has mean $R$. A simple cycle through $q_C$ that exits on $D$ has mean
\[
b_N=\frac{NP+T}{N+1}.
\]
Every other simple cycle resets at a defection state: its rewards consist of zero or more copies of $P$ and one $S$, so its mean is at most $P$. The maximum response payoff is therefore $\max\{R,b_N\}=R$, since $g(N)\ge0$ is equivalent to $b_N\le R$. Any response with at least $N+1$ states consequently has utility at most that of the proposed machine.

Now consider a response with at most $N$ states. It cannot play $C$ when the opponent first reaches $q_C$, or on any later visit to $q_C$. Such a visit is preceded by $N$ consecutive rounds in which both machines output $D$. The response would have to output $D$ on these $N$ identical inputs and then output $C$. Its states at these $N+1$ times must be distinct: a repetition among the first $N$ states would trap it in a deterministic $D$-output cycle under input $D$, and the final $C$-output state cannot equal any of them. This contradicts its size. Thus the cooperative self-loop is unavailable to a smaller response. Its average payoff is at most $b_N$, attained by $M_D$, and its utility is at most $b_N-\lambda$. The pair is therefore an NE whenever
\[
R-\lambda(N+1)\ge b_N-\lambda
\quad\Longleftrightarrow\quad\lambda\le g(N).
\]
Choosing an integer maximizing $g$ proves sufficiency up to the stated threshold.

\emph{Necessity.} Consider an NE yielding $(R,R)$ at $\lambda>0$. By Lemma~\ref{lem:structure}, both machines have the same size $n$, and by Lemma~\ref{lem:one}, $n\ge2$. Because $(C,C)$ uniquely maximizes the sum of stage payoffs, every outcome in the eventual cycle is $(C,C)$.

We use Theorem~$1^*$ of \citet{abreu1988structure}, for the limit-of-means criterion. It states that each machine's cycle states are distinct along one joint cycle; its noncycle states occur once at the beginning; and, outside a possible intermediate phase using cycle states, the players' actions correspond one-to-one. The theorem applies directly: their preferences depend on game payoff and machine size, increase with payoff at fixed size, and decrease with size at fixed payoff. For $\lambda>0$, our utility $\pi_i-\lambda|M_i|$ satisfies all three assumptions. Their limit-of-means payoff and Moore-machine representation also agree with ours.

Each machine has only one cycle state. Indeed, if it had several, a player could retain its introductory states and replace its cooperative cycle states by a single state outputting $C$ forever. Against the unchanged opponent, the introductory play and eventual payoff $R$ would be preserved, with fewer states. With only one cycle state per machine, the intermediate phase is empty: once both machines are in cycle states, their joint cycle has begun. The action correspondence must match $C$ with $C$, since that pair occurs in the cycle; consequently it matches $D$ with $D$ in the introductory phase.

All states are visited on the equilibrium path, since deleting unvisited states would save cost without changing play. Every $D$-output state is therefore introductory, and its transition on input $D$ goes to the next state in that finite introductory order. These transitions cannot form an all-$D$ cycle. Against the constant defector, every eventual cycle of the opponent must consequently contain a $C$-output state. Its length is at most $n$, so $M_D$ earns at least
\[
P+\frac{T-P}{n}.
\]
The equilibrium condition gives
\[
R-\lambda n\ge P+\frac{T-P}{n}-\lambda,
\qquad
\lambda\le\frac{(n-1)(R-P)-(T-R)}{n(n-1)}=g(n-1).
\]
This proves the upper bound. Notice that the argument does not require every introductory state to output $D$; it only needs the stated restriction on transitions from $D$-output states.
\end{proof}

\begin{lemma}[Choosing the password length]\label{lem:password-length}
Let
\[
x=\frac{T-R}{R-P}+\sqrt{\left(\frac{T-R}{R-P}\right)^2+\frac{T-R}{R-P}}.
\]
An integer maximizing $g(N)$ lies in $\{\max(1,\lfloor x\rfloor),\max(1,\lceil x\rceil)\}$.
\end{lemma}
\begin{proof}
For real $z>0$, the sign of $g'(z)$ is the sign of
\[
-(R-P)z^2+2(T-R)z+(T-R).
\]
This polynomial has exactly one positive root, $x$. Hence $g$ increases up to $x$ and decreases thereafter. Its maximum on the positive integers must be attained at an adjacent integer, or at $1$ if $x<1$.
\end{proof}

\subsection{Semi-perfect equilibrium and alternating exploitation}
\begin{proposition}[Mutual cooperation is not an SPE]\label{prop:no-coop-spe}
For every $\lambda>0$, no semi-perfect equilibrium yields the payoff $(R,R)$.
\end{proposition}
\begin{proof}
Any pair yielding $(R,R)$ eventually plays $(C,C)$ forever. From a continuation in that cycle, replacing a multi-state machine by $M_C$ produces exactly the same future actions against the unchanged opponent, preserving payoff $R$ and reducing cost. If the pair were an SPE, both machines would therefore have to use one state. But $(M_C,M_C)$ is not an NE: switching to $M_D$ increases payoff from $R$ to $T$ at the same cost. Both possibilities contradict SPE.
\end{proof}

\begin{proposition}[Upper bound on $\lambda_3$]\label{prop:spe-upper}
For $\lambda>0$ with $\lambda>(T+S)/2-P$, the unique semi-perfect equilibrium is $(M_D,M_D)$. Hence
\[
\lambda_3\le\max\{0,(T+S)/2-P\}.
\]
\end{proposition}
\begin{proof}
By Lemma~\ref{lem:structure}, any positive-cost SPE is also a lexicographic SPE. The necessity argument in \citet[Propositions~1--2]{rubinstein1986finite} restricts a non-defection SPE cycle to $(C,D)$ and $(D,C)$. Although that paper states the Prisoner's Dilemma result for numerical payoffs, this part of its argument uses only strict dominance of $D$ and the ordering $T>R>P>S$: cycle actions must lie on one diagonal, and a cycle on the $(C,C),(D,D)$ diagonal containing cooperation admits a profitable deviation. Thus the same restriction holds for our payoff parameters. Its total game payoff is therefore $T+S$. Each machine has at least two states, while either player can guarantee $P-\lambda$ by switching to $M_D$. Thus
\[
2P-2\lambda\le U_1+U_2\le T+S-4\lambda,
\]
which requires $\lambda\le(T+S)/2-P$. Above this bound at positive cost, only $(M_D,M_D)$ can remain, and that pair is an SPE at every cost. If $T+S\le2P$, no non-defection SPE survives any positive cost, so $\lambda_3=0$; this explains the maximum with zero.
\end{proof}

To obtain lower bounds, consider the two-state alternator $M^a$, with $\sigma(c)=C$, $\sigma(d)=D$, and
\[
\mu(c,\cdot)=d,\qquad\mu(d,C)=c,\qquad\mu(d,D)=d.
\]
Player 1 starts at $c$ and Player 2 at $d$.
The pair $(M^a(c),M^a(d))$ alternates between $(C,D)$ and $(D,C)$, earning each player $(T+S)/2$ and visiting both states every cycle.

\begin{proposition}[Survival threshold of the alternator]\label{prop:alt}
For $\lambda>0$, the pair $(M^a(c),M^a(d))$ is a Nash equilibrium if and only if
\[
\lambda\le\lambda_{\mathrm{alt}}:=\min\left\{\frac{T-R}{2},\frac{T+S}{2}-P\right\}.
\]
Whenever it is an NE, it is also an SPE. In the regular regime $2P\le R+S$, this threshold is $(T-R)/2$.
\end{proposition}
\begin{proof}
Fix either initial state of the opponent. Both states are reachable in its response graph. The graph has edges
\[
d\xrightarrow{C,S}c,\quad d\xrightarrow{D,P}d,
\qquad c\xrightarrow{C,R}d,\quad c\xrightarrow{D,T}d,
\]
where each edge is labelled by the responding player's action and payoff. Its simple-cycle means are $P$, $(R+S)/2$, and $(T+S)/2$. If $(T+S)/2\le P$, switching from the two-state alternator to $M_D$ improves utility at every positive cost, so no such equilibrium exists.

Otherwise the maximum cycle mean is $(T+S)/2$, attained by the two-state response. Lemma~\ref{lem:fixed} then bounds the utility of every response with at least two states by $(T+S)/2-2\lambda$. The only one-state responses are $M_D$ and $M_C$, earning respectively $P$ and $(R+S)/2$. Neither is profitable precisely when
\[
\frac{T+S}{2}-2\lambda\ge P-\lambda,
\qquad
\frac{T+S}{2}-2\lambda\ge\frac{R+S}{2}-\lambda.
\]
These inequalities are equivalent to the displayed bound on $\lambda$. The calculation applies from both $c$ and $d$. Since the joint play visits only $(c,d)$ and $(d,c)$, it proves SPE whenever it proves NE. Finally, $(T-R)/2\le(T+S)/2-P$ holds exactly when $2P\le R+S$.
\end{proof}

\begin{proposition}[Lower bounds]\label{prop:lower}
If $T+S>2P$, then $\lambda_3\ge\lambda_{\mathrm{alt}}$ and $\lambda_4\ge\max\{\lambda_2,\lambda_{\mathrm{alt}}\}$.
\end{proposition}
\begin{proof}
When $T+S>2P$, the alternator has a payoff different from $(P,P)$ and $\lambda_{\mathrm{alt}}>0$. Proposition~\ref{prop:alt} supplies a non-defection SPE, and hence NE, through that cost. Proposition~\ref{prop:lambda2} supplies an NE yielding $(R,R)$ through $\lambda_2$. Taking the suprema in the definitions gives both bounds.
\end{proof}

Thus the proven bounds in the regular regime are
\[
\lambda_1=0,\quad
\lambda_{\mathrm{alt}}\le\lambda_3\le(T+S)/2-P,\quad
\max\{\lambda_2,\lambda_{\mathrm{alt}}\}\le\lambda_4\le R-P.
\]
The two constructions motivate the conjectures below; their optimality for $\lambda_3$ and $\lambda_4$ remains open.

\section{Computational Evidence and Future Directions}
Preliminary computational testing, exhaustive over machine
pairs with at most two states and supplemented by seeded
searches up to three states, corroborates the conjectures
$\lambda_3=\lambda_{\mathrm{alt}}$ and
$\lambda_4=\max\{\lambda_2,\lambda_{\mathrm{alt}}\}$
across five tested payoff parametrizations satisfying
$2P\le R+S$. The tested cases include both
$\lambda_2<\lambda_{\mathrm{alt}}$ and
$\lambda_2>\lambda_{\mathrm{alt}}$, suggesting that eventual
mutual cooperation and alternating exploitation need not
disappear in a fixed order. Establishing these conjectures
for machines of arbitrary size remains open. A further
direction is to use these thresholds in resource-aware
synthesis: constructing finite-state strategies that satisfy
an eventual-cooperation specification while remaining stable
against cost-sensitive replacement.

\bibliographystyle{plainnat}
\bibliography{references}

@article{veelen2019,
  title={In and out of equilibrium II: Evolution in repeated games with discounting and complexity costs},
  author={van Veelen, Matthijs and Garc{\'\i}a, Juli{\'a}n},
  journal={Games and Economic Behavior},
  volume={115},
  pages={113--130},
  year={2019},
  publisher={Elsevier}
}

@article{aumann1981survey,
  title={Survey of repeated games},
  author={Aumann, Robert J},
  journal={Essays in game theory and mathematical economics in honor of Oskar Morgenstern},
  year={1981},
  publisher={Wissenschaftsverlag Bibliographisches Institut}
}

@article{neyman1985bounded,
  title={Bounded complexity justifies cooperation in the finitely repeated prisoners' dilemma},
  author={Neyman, Abraham},
  journal={Economics letters},
  volume={19},
  number={3},
  pages={227--229},
  year={1985},
  publisher={Elsevier}
}

@article{abreu1988structure,
  title={The structure of Nash equilibrium in repeated games with finite automata},
  author={Abreu, Dilip and Rubinstein, Ariel},
  journal={Econometrica: Journal of the Econometric Society},
  pages={1259--1281},
  year={1988},
  publisher={JSTOR}
}

@article{rubinstein1986finite,
  title={Finite automata play the repeated prisoner's dilemma},
  author={Rubinstein, Ariel},
  journal={Journal of economic theory},
  volume={39},
  number={1},
  pages={83--96},
  year={1986},
  publisher={Elsevier}
}

@article{binmore1990,
  title={Evolutionary stability in repeated games played by finite automata},
  author={Binmore, Kenneth G and Samuelson, Larry},
  journal={Journal of economic theory},
  volume={57},
  number={2},
  pages={278--305},
  year={1992},
  publisher={Elsevier}
}

@inproceedings{megiddo1986play,
  title={On play by means of computing machines: preliminary version},
  author={Megiddo, Nimrod and Wigderson, Avi},
  booktitle={Theoretical aspects of reasoning about knowledge},
  pages={259--274},
  year={1986},
  organization={Elsevier}
}

@article{friedman1971non,
  title={A non-cooperative equilibrium for supergames},
  author={Friedman, James W},
  journal={The Review of Economic Studies},
  volume={38},
  number={1},
  pages={1--12},
  year={1971},
  publisher={Wiley-Blackwell}
}

@article{bankssundaram1990,
  title={Repeated games, finite automata, and complexity},
  author={Banks, Jeffrey S and Sundaram, Rangarajan K},
  journal={Games and Economic Behavior},
  volume={2},
  number={2},
  pages={97--117},
  year={1990},
  publisher={Elsevier}
}

@article{cooper1996evo,
  title={Supergames played by finite automata with finite costs of complexity in an evolutionary setting},
  author={Cooper, David J},
  journal={Journal of Economic Theory},
  volume={68},
  number={1},
  pages={266--275},
  year={1996},
  publisher={Elsevier}
}
\end{document}